\documentclass[10pt,twocolumn]{article}
\usepackage[T1]{fontenc}
\usepackage{graphicx,amsmath,amssymb,bm,booktabs,cite,microtype,dblfloatfix,balance,float}
\usepackage[margin=1.78cm,top=1.65cm,bottom=1.75cm,columnsep=0.58cm]{geometry}
\usepackage[hidelinks]{hyperref}
\graphicspath{{figures/}}
\newcommand{\snn}{\sqrt{s_{\mathrm{NN}}}}
\newcommand{\GeVfm}{\mathrm{GeV/fm}^{3}}

\newcommand{\vtwo}{v_{2}}

\begin{document}

\twocolumn[
\begin{@twocolumnfalse}
\begin{center}
{\Large\bfseries Strange-quark-number-dependent response to partonic dynamics\\
in baryon-rich Au+Au collisions}\par
\vspace{0.55em}
{\large Di Zhao$^{1,2}$, Xun Zhu$^{1,2}$, and Gao-Chan Yong$^{1,2,3}$}\par
\vspace{0.35em}
{\small
$^1$Institute of Modern Physics, Chinese Academy of Sciences, Lanzhou 730000, China\\
$^2$School of Nuclear Science and Technology, University of Chinese Academy of Sciences, Beijing 100049, China\\
$^3$State Key Laboratory of Heavy Ion Science and Technology, Institute of Modern Physics, Chinese Academy of Sciences, Lanzhou 730000, China}
\end{center}
\vspace{0.4em}
\begin{abstract}
We identify a strange-quark-number-dependent hierarchy in the response to explicit partonic dynamics by comparing PHSD5.2 with its hadron--string limit in baryon-rich Au+Au collisions at $\snn=3.0$--$7.2$~GeV. Crossing the local conversion threshold alone does not ensure a visible final-state signal: abundant-hadron flow changes little at 3--4.5~GeV, whereas the fractional yield response becomes ordered at 6.2--7.2~GeV, reaching $S_{|s|=1}\simeq0.2$--$0.3$, $S_{|s|=2}\simeq0.65$, and $S_{|s|=3}\simeq0.93$ at 7.2~GeV. The hierarchy is consistent with an enlarged recombination phase space for strange constituents released from different pre-hadronic sources, but does not by itself establish an increase in the total $s+\bar s$ abundance or stronger strange-quark clustering. Since $S_X$ is a model-defined response rather than a direct observable, matched excitation functions of $\Xi^-/\Lambda$ and $\Omega^-/\Lambda$, together with high-$m_T$ proton elliptic flow, provide experimentally accessible discriminants between the predicted PHSD and HSD dynamics.
\end{abstract}
\vspace{0.55em}
\end{@twocolumnfalse}
]

\section{Introduction}

The phase structure of quantum chromodynamics at large baryon chemical potential remains poorly constrained. Lattice QCD establishes an analytic crossover at vanishing chemical potential, but the sign problem prevents a direct determination of the high-density phase boundary \cite{Fukushima2011,Borsanyi2020,Bzdak2020}. Beam-energy-scan and fixed-target programs therefore seek observables that retain an early-stage imprint of a change from hadronic to partonic degrees of freedom \cite{Luo2020}. At $\snn=7.7$--$62.4$~GeV, STAR found approximate constituent-quark-number scaling of the elliptic flow for a selected group of identified hadrons, within about 10\% in the specified kinematic range, providing strong evidence that partonic collectivity contributes in this energy domain \cite{STARBESv2}. This evidence is not an isolated proof of a thermodynamic phase transition, particularly because particle--antiparticle flow splitting grows toward the lower energies. Current measurements below 7.7~GeV bracket a particularly interesting interval: collectivity at $3$~GeV is predominantly hadronic, whereas the gradual recovery of constituent-quark-number scaling toward $4.5$~GeV indicates that partonic interactions may become relevant \cite{STAR3GeV,STAR3GeVErratum,STARonset}. These observations motivate searching for a change of effective degrees of freedom between about 3 and 7.7~GeV. The present study targets precisely the $\snn=3.0$--$7.2$~GeV interval.

The central difficulty is that final hadrons integrate compression, mean fields, spectator shadowing, resonance and string dynamics, hadronization, and rescattering. Flow is sensitive to early pressure but also to hadronic potentials, the equation of state, and spectators \cite{Brachmann2000,Nara2022}. Strangeness production was proposed as a diagnostic of a deconfined medium \cite{RafelskiMuller1982}, but strange hadrons can also be generated by string fragmentation, chiral-symmetry-restoration effects, in-medium reactions, and hadronic strangeness exchange without deconfinement \cite{CassingCSR,Palmese2016,Song2021,Pal2004}. Purely hadronic transport, for example, predicts a stronger equation-of-state response for $\Xi^-$ than for singly strange particles because its production is more strongly multistep \cite{YongXi2021}. A hierarchy with strange content is therefore not unique to partonic matter; a useful signal must be tested against these hadronic alternatives.

Previous AMPT calculations found that parton scattering modifies strange-quark and strange-baryon distributions below $\snn\sim10$~GeV and proposed a momentum-dependent single-to-double-strangeness ratio as a quark-matter discriminator \cite{YongQM2023}. Here we identify a complementary pattern within PHSD: the fractional response to explicit partonic dynamics increases systematically with the number $N_s$ of valence strange or antistrange constituents in the observed hadron. The result follows from a controlled comparison of PHSD5.2 with its Hadron--String Dynamics (HSD) limit. The large responses of $\Xi^-(dss)$ and $\Omega^-(sss)$ may reflect altered recombination opportunities, additional partonic production, or different production and survival histories. The final yields alone do not distinguish these mechanisms. We therefore contrast the model-response hierarchy with proton elliptic flow and formulate experimentally accessible consequences.

\section{Framework and observables}

PHSD is a covariant off-shell transport approach derived from the Kadanoff--Baym equations \cite{Cassing2008,Cassing2009}. Its partonic sector employs massive quarks, antiquarks, and gluons with finite spectral widths and interactions constrained by lattice-QCD thermodynamics. Color-neutral matter is formed through covariant fusion of $q\bar q$ pairs and three (anti)quarks, conserving four-momentum and flavor currents \cite{Bratkovskaya2011,Moreau2019}. The corresponding HSD calculation retains the same hadronic, resonance, string, and chiral-symmetry-restoration dynamics but contains no explicit partonic stage. Their difference therefore measures the response to the additional partonic degrees of freedom within this common transport framework.

In the employed PHSD5.2 setup, eligible pre-hadronic matter is dissolved when the local energy density exceeds $\varepsilon_c=0.4\ \GeVfm$; hadronization is admitted below $0.65\ \GeVfm$. These numerical cuts define an operational conversion window, not a universal equilibrium phase boundary. A cell above threshold is only a necessary condition. Even if conversion occurs, a small partonic fraction or short partonic lifetime may leave no resolvable imprint after hadronization and hadronic rescattering.

We study Au+Au collisions at $\snn=3.0$, 4.5, 5.2, 6.2, and 7.2~GeV. Flow coefficients are $v_n=\langle\cos[n(\phi-\Psi_{\rm RP})]\rangle$ \cite{Poskanzer1998}. Strange-particle yields are inclusive full-phase-space multiplicities in 10--30\% central collisions. For species $X$ we define
\begin{equation}
 S_X=\frac{N_X^{\rm PHSD}-N_X^{\rm HSD}}{N_X^{\rm HSD}}.
 \label{eq:sens}
\end{equation}
This ratio removes the dominant excitation-function growth and exposes the response to switching on the explicit partonic sector within PHSD. It is a model-defined counterfactual quantity: experiment cannot construct $S_X$ directly, but can compare measured absolute yields and particle ratios with the separate PHSD and HSD predictions.

\section{Results}

\begin{figure}[t!]
\centering
\includegraphics[width=0.9\columnwidth]{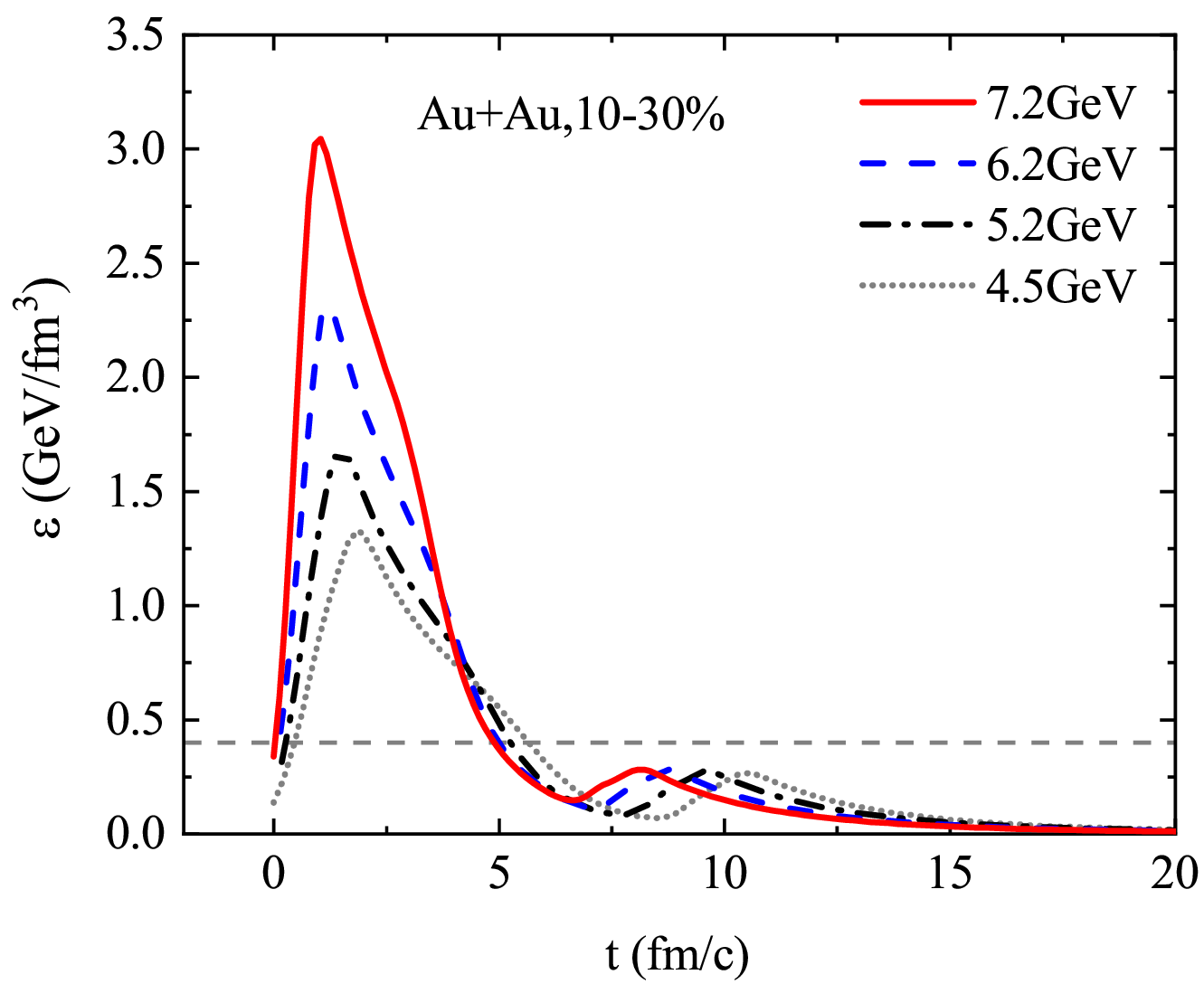}
\caption{Time evolution of the energy density in the central cell ($x=y=z=0$) of 10--30\% Au+Au collisions. The horizontal dashed line denotes the critical energy density $\varepsilon_c=0.4\ \GeVfm$ used in the calculation.}
\label{fig:edens}
\vspace{0.5cm}
\end{figure}
We begin by determining whether the compressed reaction zone reaches the operational energy-density window in which PHSD permits the conversion of pre-hadronic matter into partonic degrees of freedom. For this purpose, the central cell provides a direct time-resolved diagnostic of the maximum compression region. It does not represent the full reaction volume and is not itself an observable; rather, it supplies the necessary dynamical reference for interpreting whether differences in the subsequent hadronic observables can plausibly originate from a partonic stage.

Fig.~\ref{fig:edens} shows that the energy density of the central cell exceeds $\varepsilon_c$ already at 4.5~GeV. Its peak rises from approximately $1.3\ \GeVfm$ at 4.5~GeV to $3.0\ \GeVfm$ at 7.2~GeV. Nevertheless, threshold crossing does not imply observable sensitivity. A small space--time-integrated partonic fraction or a short partonic lifetime could prevent an early difference from surviving hadronization and late rescattering. The central-cell curve alone, however, determines neither quantity, so the final-state response must be assessed independently.
The central-cell evolution therefore identifies when partonic conversion becomes possible in the calculation, whereas the survival of its imprint must be assessed independently with final-state flow and particle-production observables.

\begin{figure}[t!]
\centering
\includegraphics[width=0.9\columnwidth]{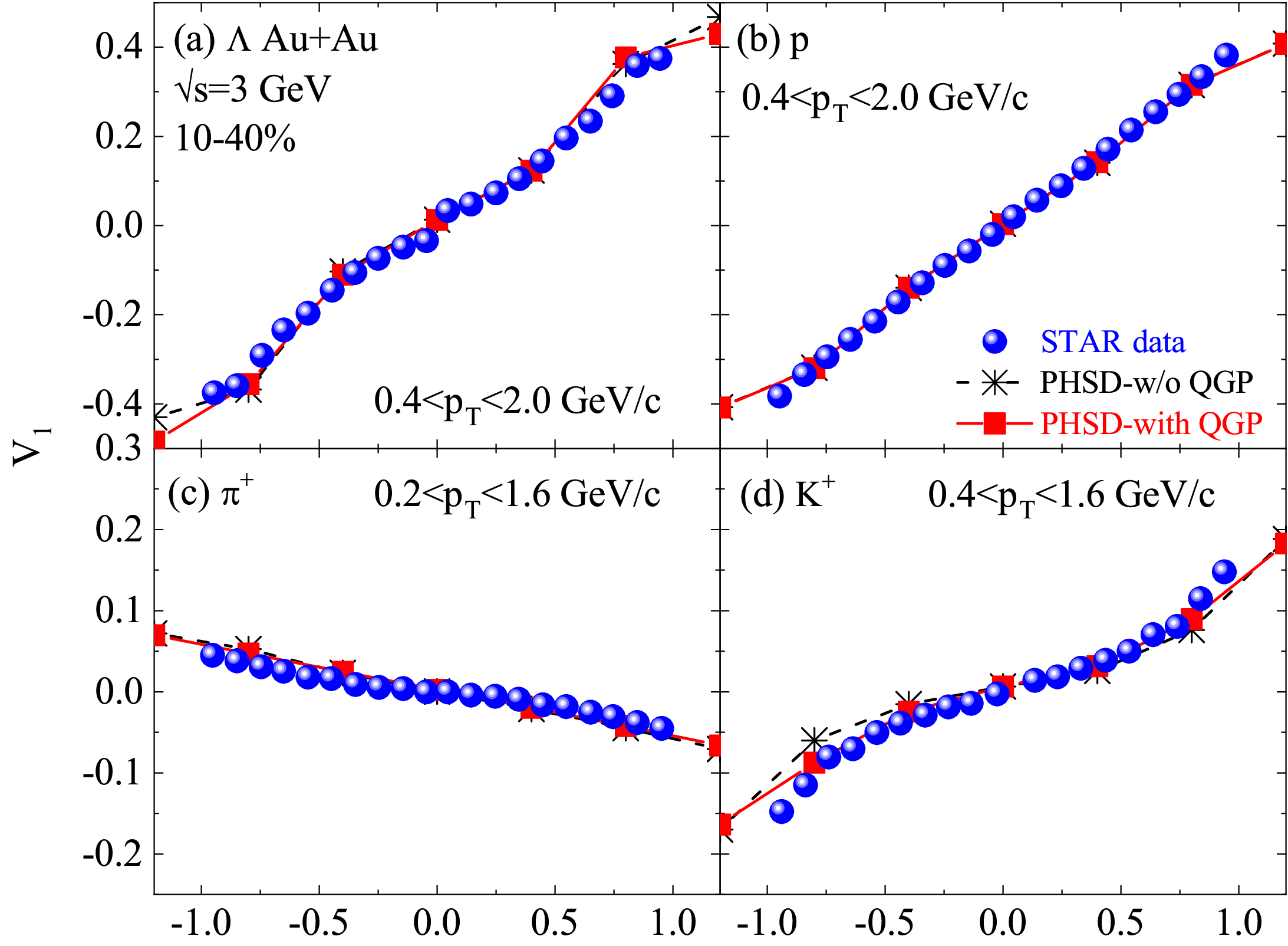}\\[-0.25em]
\includegraphics[width=0.9\columnwidth]{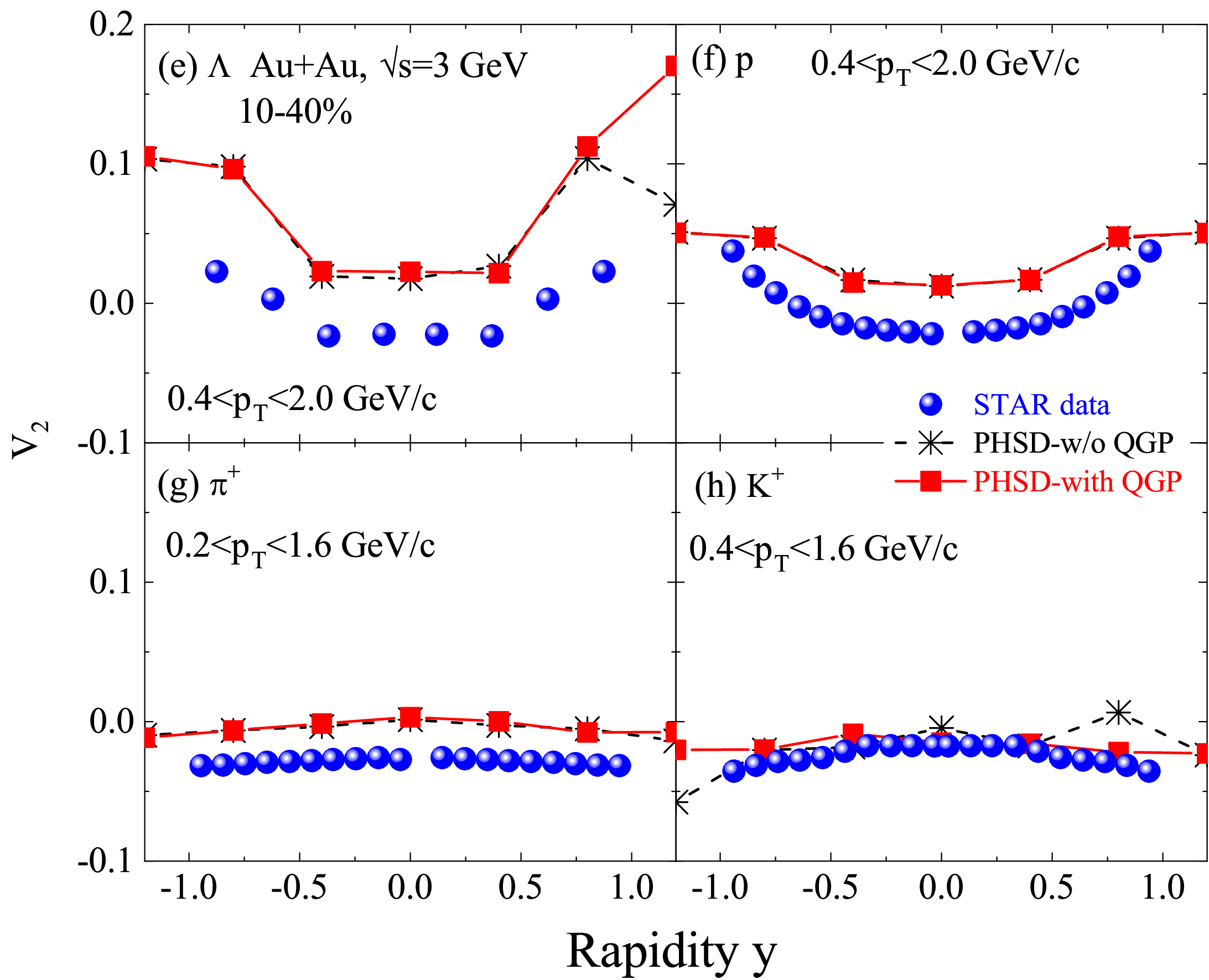}
\caption{Rapidity dependence of (a--d) directed and (e--h) elliptic flow for $\Lambda$, $p$, $\pi^+$, and $K^+$ in 10--40\% Au+Au collisions at $\snn=3$~GeV. Red solid lines denote PHSD, black dashed lines HSD, and blue circles STAR data \cite{STAR3GeV}.}
\label{fig:flow3}
\vspace{0.5cm}
\end{figure}
With this temporal benchmark established, we first examine whether the abundant identified hadrons retain a coherent response at the lowest beam energy. Directed flow reflects the sideward collective deflection, while elliptic flow at these energies results from the competition between pressure-driven expansion and spectator shadowing \cite{Brachmann2000,Nara2022}. Considering both coefficients for several species tests whether the activation of partonic degrees of freedom produces a common change rather than an isolated species-dependent variation.

At 3~GeV the PHSD and HSD curves in Fig.~\ref{fig:flow3} are nearly indistinguishable. Both modes reproduce much of the directed-flow ordering but underestimate the measured negative elliptic flow. Because both modes share the same limitation, this discrepancy cannot be attributed to the partonic switch; low-energy baryonic dynamics, potentials, cluster formation, and spectator shadowing are possible sources that require separate study. The same near-degeneracy persists for the calculated abundant-hadron directed flow at 4.5~GeV. Thus, even when the central-cell density is well above $\varepsilon_c$, low-order flow of abundant hadrons is not a clean stand-alone discriminator in this setup.

Momentum-integrated bulk flow can dilute an early difference, so we next focus on the transverse-kinetic-energy dependence. Protons offer sufficient abundance for this differential comparison throughout the energy scan, and the high-$m_T-m_0$ region preferentially samples particles that acquired substantial momentum during the earlier, more compressed stage. Proton elliptic flow is therefore used here as a dynamical cross-check of the PHSD--HSD separation, not as direct evidence for deconfinement by itself.
\begin{figure}[t!]
\centering
\includegraphics[width=0.98\columnwidth]{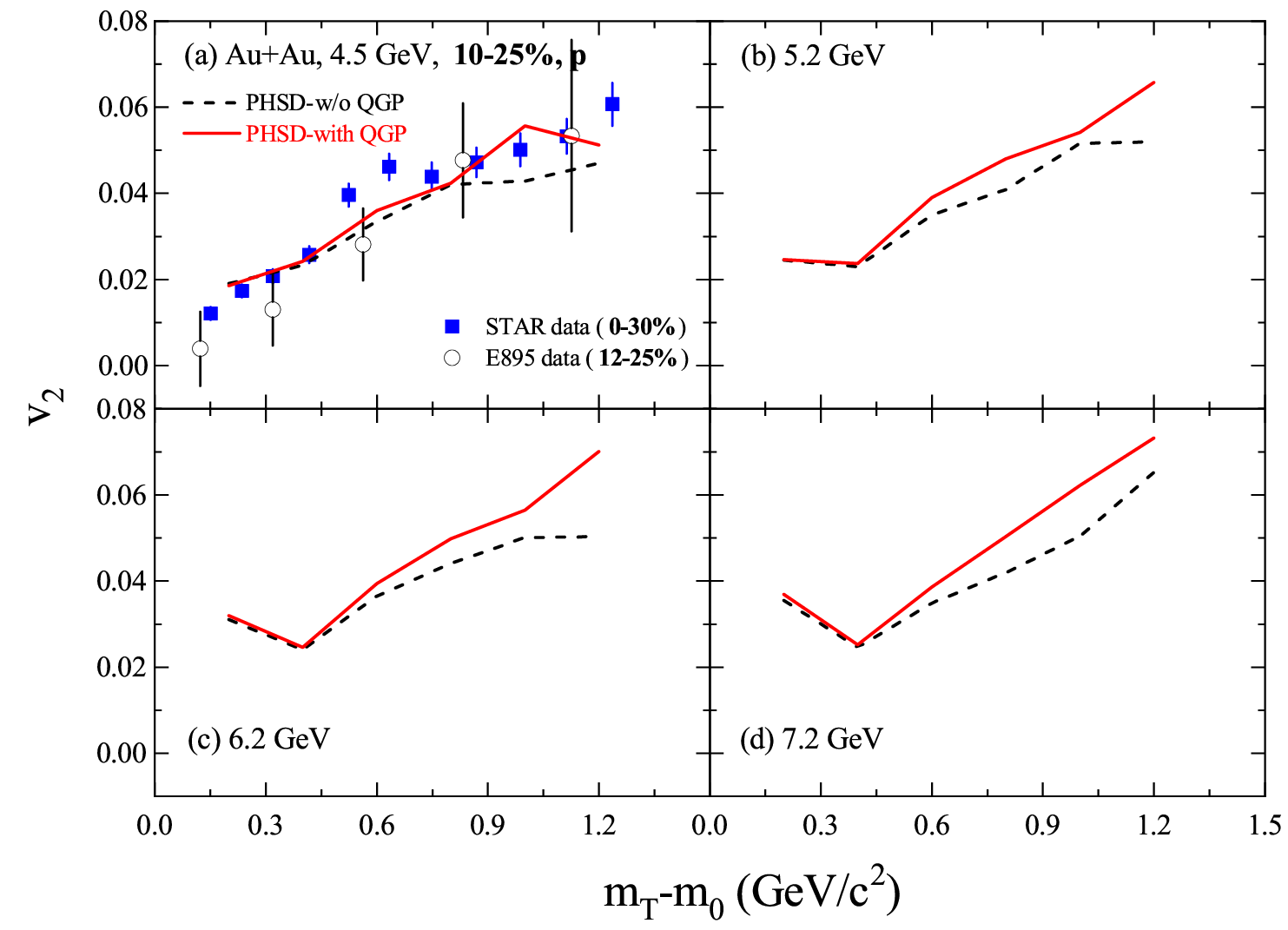}
\caption{Proton $\vtwo$ versus transverse kinetic energy in midcentral Au+Au collisions. PHSD (red solid) develops a moderate separation from HSD (black dashed) at large $m_T-m_0$. Data at 4.5~GeV are from STAR \cite{STAR45} and E895 \cite{E895v2}.}
\label{fig:v2}
\vspace{0.25cm}
\end{figure}
As shown in Fig.~\ref{fig:v2}, the two modes remain close at small transverse kinetic energy, whereas PHSD generally produces larger $\vtwo$ at high $m_T-m_0$. The high-$m_T-m_0$ region can be more responsive to early pressure, but it is not an exclusive early-time tag and remains entangled with baryonic mean fields and spectator blocking \cite{Nara2022}. It is therefore useful as an independent consistency test rather than as the primary chemical signature.

Flow alone does not isolate the flavor content of the additional response. We therefore turn to particle production across the energy scan and compare hyperons in the same centrality class. The sequence $\Lambda,\Sigma^0\rightarrow\Xi^-\rightarrow\Omega^-$ increases the number of strange valence quarks from one to three, so the corresponding excitation functions can reveal whether the PHSD--HSD separation develops an ordered dependence on hadron quark content.
\begin{figure}[t!]
\centering
\includegraphics[width=0.98\columnwidth]{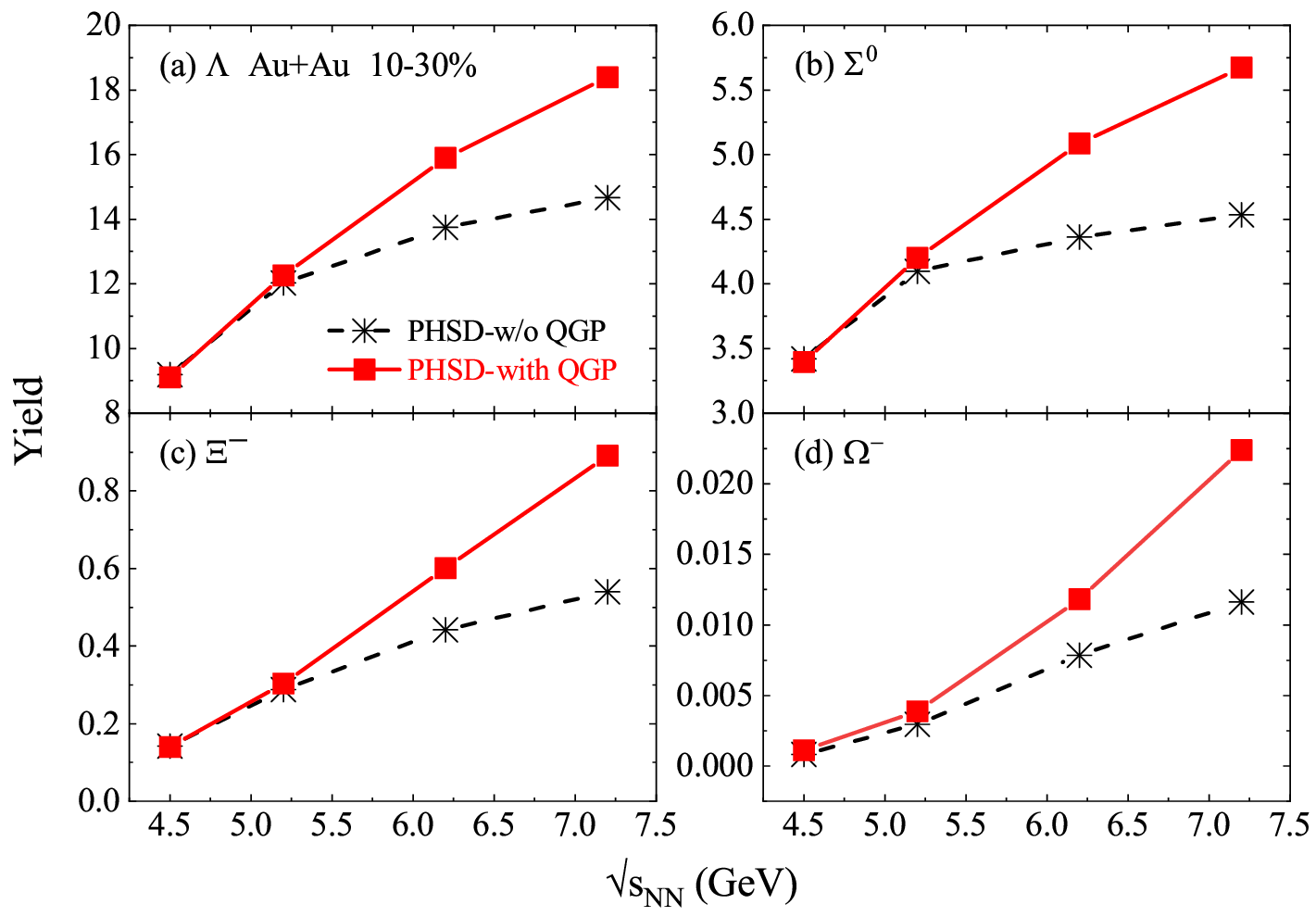}
\caption{Excitation functions of hyperon yields in 10--30\% Au+Au collisions. The PHSD--HSD separation grows with energy and strange valence-quark number.}
\label{fig:yields}
\vspace{0.5cm}
\end{figure}
As shown in Fig.~\ref{fig:yields}, the hyperon excitation functions reveal the clearest ordered response. At 4.5~GeV the two modes are nearly identical, apart from the statistically fragile fractional change of the very rare $\Omega^-$. A qualitative ordering begins at 6.2~GeV and is clearest at 7.2~GeV: the separation is modest for singly strange $\Lambda$ and $\Sigma^0$, larger for $\Xi^-$, and largest for $\Omega^-$. Normalization by the HSD yield makes this hierarchy more visible, but also amplifies fluctuations when the baseline is small. Thus, the relevant information is not one large rare-channel ratio in isolation, but a coherent ordering of absolute and fractional PHSD--HSD differences across species and neighboring beam energies.

Absolute yields still combine the overall growth of particle production with the species-dependent response. To compare unlike strange hadrons on the same scale, we next recast the calculated differences in terms of the fractional sensitivity $S_X$ defined in Eq.~(\ref{eq:sens}) and include kaons as one-strange mesonic references. This normalization tests whether the observed ordering follows absolute valence strangeness rather than simply the magnitude of a species yield.
\begin{figure}[t!]
\centering
%\begin{center}
\includegraphics[width=0.98\columnwidth]{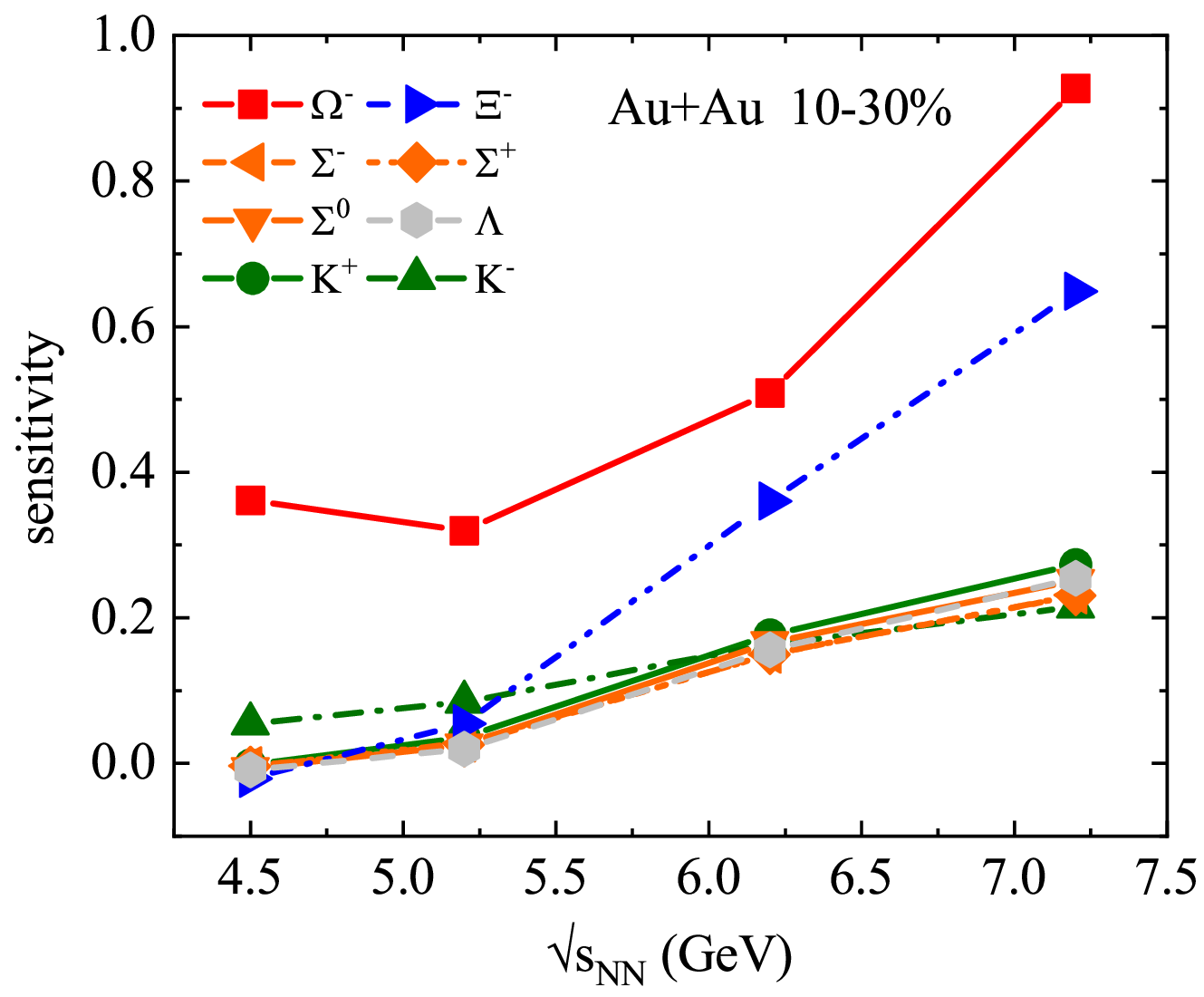}
\caption{Fractional partonic response $S_X$ [Eq.~(\ref{eq:sens})]. A qualitative ordering by absolute valence strangeness begins at 6.2~GeV and is clearest at 7.2~GeV.}
\label{fig:ratio}
%\end{center}
\vspace{0.5cm}
\end{figure}
At 7.2~GeV, Fig.~\ref{fig:ratio} gives $S_X\simeq0.21$--$0.27$ for kaons and singly strange hyperons, $S_{\Xi^-}\simeq0.65$, and $S_{\Omega^-}\simeq0.93$. The important new result is the correlated ordering
\begin{equation}
 S_{N_s=3}>S_{N_s=2}>S_{N_s=1},
 \label{eq:ordering}
\end{equation}
rather than any isolated enhancement.

Here ``sensitivity'' denotes the fractional model response, not experimental resolving power. In particular, the large $S_{\Omega^-}$ can partly reflect the small HSD denominator. A reliable hierarchy requires the absolute PHSD--HSD differences to exceed their statistical uncertainties and the ordering to persist across adjacent energies and reasonable model variations. Accordingly, the proposed result is the cross-species ordering itself, not an exact universal proportionality between $S_X$ and $N_s$.

The ordering admits a coalescence interpretation, but the final yields do not determine its microscopic origin uniquely. In a schematic coalescence counting argument, rather than an exact PHSD yield formula, the leading flavor dependence can be represented as \cite{Greco2003}
\begin{equation}
 N_h\propto \int d\Gamma\,\rho_s^{\,N_s}\rho_q^{\,N_q}\,W_h,
 \label{eq:coal}
\end{equation}
where $\rho_s$ denotes the appropriate local $s$- or $\bar s$-phase-space density, $\rho_q$ represents the light-flavor density, $N_s$ counts valence strange or antistrange constituents, and $W_h$ contains overlap, conservation, and dynamical factors. Eq.~(\ref{eq:coal}) depends on higher local density moments and on the recombination kernel, not only on the integrated $s+\bar s$ abundance. In PHSD, dissolution places constituent quarks from pre-hadronic sources into the partonic transport sector, after which flavor-conserving fusion can in principle combine quarks with different parentage \cite{Bratkovskaya2011,Moreau2019}. This mechanism can change the accessible recombination phase space without necessarily increasing the total $s+\bar s$ abundance or producing genuine spatial clustering. The present final-yield analysis contains no quark-parentage information and therefore does not establish that this mechanism causes the hierarchy.

Only under the additional assumption that the relevant local strange-quark density changes approximately as $\rho_s\rightarrow(1+\delta_s)\rho_s$, while other factors largely cancel, does Eq.~(\ref{eq:coal}) reduce to
\begin{equation}
 S(N_s)\simeq(1+\delta_s)^{N_s}-1\simeq N_s\delta_s
 \quad (\delta_s\ll1).
 \label{eq:scaling}
\end{equation}
Eq.~(\ref{eq:scaling}) is therefore a conditional organizing estimate, not a demonstrated scaling law or an extraction of $\delta_s$. Exact proportionality is neither expected nor claimed: hadronic production and absorption, spatial correlations, quark rearrangement, and species-dependent kernels can all modify the response. Moreover, only a qualitative ordering begins at 6.2~GeV and it becomes pronounced at 7.2~GeV; the isolated low-energy $\Omega^-$ response is not interpreted because its HSD denominator is very small.

Several effects can make the fractional response larger for multi-strange baryons. Singly strange particles have large hadronic baselines from associated production and strings, whereas $dss$ or $sss$ formation is intrinsically more multistep in a purely hadronic description. Hadronic strangeness-exchange reactions can themselves contribute substantially to multi-strange production \cite{Pal2004,YongXi2021}; in PHSD, flavor-conserving fusion additionally permits a rearrangement of partonic constituents. The present calculation does not separate these contributions from additional production, absorption, or survival effects. Conversely, the low $\Omega^-$ yield demands high statistics and makes its fractional response especially vulnerable to a small denominator. Determining the dominant mechanism requires a time-resolved inventory of $s+\bar s$, separation of production, dissolution, fusion, and absorption channels, and parentage information for the strange quarks entering final $\Xi$ and $\Omega$ baryons.

Experiment cannot measure Eq.~(\ref{eq:sens}) directly. Its accessible consequence is instead a separation between the PHSD and HSD predictions for absolute yields and for $\Xi^-/\Lambda$ and $\Omega^-/\Lambda$ in a common centrality and acceptance; high-$m_T$ proton $\vtwo$ supplies an independent dynamical comparison. A partonic interpretation would be supported only if the data consistently favor the PHSD excitation functions for several species and flow observables within their combined experimental and model uncertainties. Quantitative discrimination requires acceptance-filtered calculations and variations of the conversion window, partonic mean field, hadronic potentials, chiral-restoration strength, and poorly known multi-strange cross sections.

The results therefore form a consistent model-response sequence. The central energy density establishes the opportunity for partonic conversion, high-$m_T$ proton elliptic flow supplies a moderate pressure-related response, and multi-strange production exhibits the largest fractional PHSD--HSD separation. None of these elements alone proves a phase transition, and the calculated hierarchy is not itself an experimental observable; the falsifiable prediction is the set of absolute excitation functions and ratios produced by the two modes.

\section{Conclusions}

PHSD5.2 predicts that the most distinctive response to switching on explicit partonic dynamics in baryon-rich Au+Au collisions is an ordering with absolute valence strangeness. Threshold crossing alone is insufficient, and abundant-hadron flow at 3--4.5~GeV remains dominated by common hadronic dynamics. A qualitative ordering begins at 6.2~GeV and becomes pronounced at 7.2~GeV, where the fractional yield response grows from approximately 20--30\% for one-strange hadrons to 65\% for $\Xi^-$ and 93\% for $\Omega^-$. The large multi-strange response is consistent with modified recombination phase space, but the final yields do not establish its microscopic origin, an increase in the total $s+\bar s$ abundance, or genuine strange-quark clustering.

The hierarchy identifies multi-strange baryons as strong fractional discriminants within the PHSD--HSD comparison, although the rare $\Omega^-$ is not automatically the most precise experimental discriminator. Since $S_X$ cannot be constructed from data, the experimentally testable predictions are the separate excitation functions of $\Xi^-/\Lambda$ and $\Omega^-/\Lambda$, together with high-$m_T$ proton elliptic flow. PHSD and HSD differ in degrees of freedom, interactions, and hadronization history rather than by one isolated parameter. The comparison therefore constrains the response to partonic dynamics within PHSD but does not alone prove deconfinement, determine the order of the QCD transition, or establish a critical point.

%\section*{Declaration of competing interest}
%The authors declare that they have no known competing financial interests or personal %relationships that could have appeared to influence the work reported in this paper.

\section*{Acknowledgements}
The author GCY thanks E. L. Bratkovskaya for providing PHSD code and helpful discussions. This work is supported by the National Natural Science Foundation of China under Grant Nos.~12275322 and 12335008.

%\section*{Data availability}
%Data will be made available on request.

% BEGIN ADDED APPENDICES
\section*{Appendix}
\appendix
\raggedbottom
\section{Strangeness-production mechanisms in PHSD}
\label{app:mechanisms}

Chiral symmetry restoration (CSR) and explicit partonic reactions are distinct mechanisms. In string fragmentation, partial CSR lowers effective quark masses and reduces the suppression of strange relative to light quark-pair production \cite{CassingCSR,Palmese2016}. This mechanism is included in both the PHSD and HSD calculations considered here. Their difference therefore cannot be interpreted as the effect of introducing CSR alone. Its accumulated contribution need not cancel, however, because the two evolutions encounter different densities and spend different times in the hadronic and string sectors.

The additional partonic stage in PHSD permits strange-pair creation and annihilation through kinematically allowed $q\bar q\leftrightarrow g$ reactions. It also changes how existing strange quarks are distributed and assembled: dissolution of pre-hadrons releases their flavor content, followed by partonic transport and flavor-conserving quark--antiquark or three-quark fusion \cite{Bratkovskaya2011,Moreau2019}. Fusion itself does not create a new strange pair. Forming a $dss$ or $sss$ baryon requires several strange constituents in the appropriate phase space, so altered recombination opportunities can produce a substantial fractional response for multi-strange baryons.

The final PHSD--HSD yield difference is thus the net result of pair production, flavor redistribution, and subsequent hadronic production and absorption. Final yields alone do not identify the dominant contribution or demonstrate an increase in total strangeness or spatial clustering. The conditional coalescence argument in Eqs.~(\ref{eq:coal}) and (\ref{eq:scaling}) should be understood within this limitation; a quantitative separation would require additional reaction-history analysis.

\section{Comparison with preliminary STAR BES-II rapidity spectra}
\label{app:besii}
\setcounter{figure}{0}
\renewcommand{\thefigure}{B.\arabic{figure}}
\renewcommand{\theHfigure}{appendixB.\arabic{figure}}

\begin{figure}[H]
\centering
\includegraphics[width=0.98\textwidth]{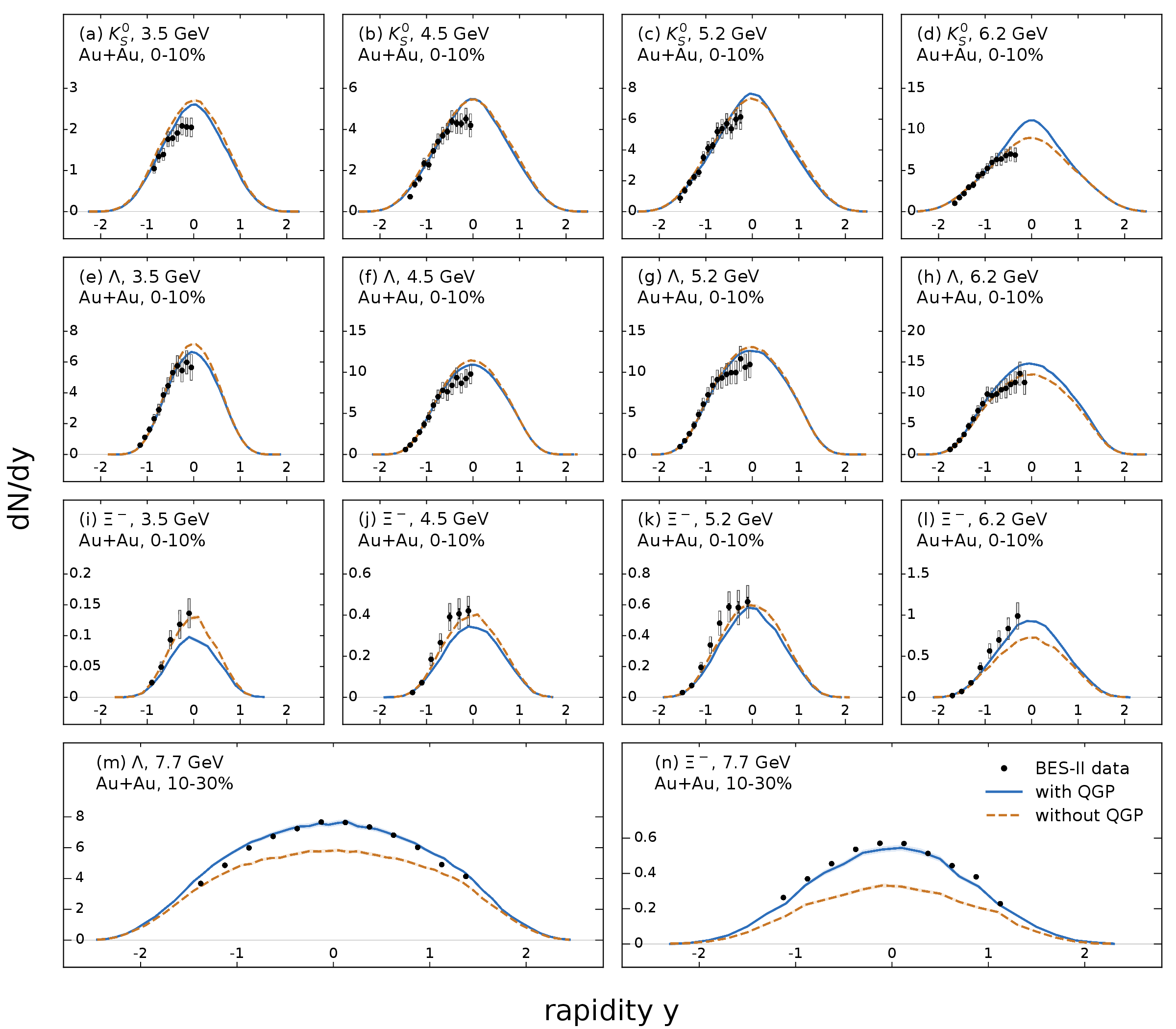}
\caption{Rapidity spectra in Au+Au collisions at 0--10\% centrality for 3.5--6.2~GeV and 10--30\% for 7.7~GeV. Black points: preliminary STAR BES-II data \cite{STARFXT2026,STARBESII2026}; the 7.7~GeV points average the two equal-width centrality intervals. Blue solid curves: PHSD with QGP; orange dashed curves: HSD without QGP. Shaded bands show model Monte Carlo standard errors. Only experimental uncertainties reliably recovered from the source plots are displayed; the available error information is incomplete, especially at 7.7~GeV.}
\label{fig:besii}
\end{figure}

Figure~\ref{fig:besii} compares PHSD with a partonic stage (``with QGP'') and its hadronic/string counterpart HSD (``without QGP'') with preliminary STAR BES-II data digitized from Ref.~\cite{STARFXT2026}, Fig.~2, and Ref.~\cite{STARBESII2026}, Fig.~1. The comparison covers $K_S^0$, $\Lambda$, and $\Xi^-$ at 3.5, 4.5, 5.2, and 6.2~GeV in 0--10\% central collisions, and $\Lambda$ and $\Xi^-$ at 7.7~GeV in 10--30\% central collisions. The cited 7.7~GeV figure provides no corresponding $K_S^0$ spectrum. Antihyperons are not compared.

For 7.7~GeV, the reported 10--20\% and 20--30\% per-event yields are averaged at each rapidity with equal weights, because these centrality intervals have equal widths. Each model calculation contains 50,100 events at 3.5 and 4.5~GeV, 30,000 at 5.2 and 6.2~GeV, and 20,100 at 7.7~GeV. Centrality is selected geometrically. The calculated $\Lambda$ yield includes electromagnetic $\Sigma^0$ decays but excludes weak hyperon feed-down.

At 3.5--5.2~GeV, the species considered do not show a uniform preference for the partonic calculation. At 6.2~GeV, it improves the $\Xi^-$ comparison, while $K_S^0$ and $\Lambda$ remain overpredicted. The clearest simultaneous improvement occurs at 7.7~GeV for $\Lambda$ and $\Xi^-$, particularly near midrapidity, although the partonic calculation overshoots some outer-rapidity $\Lambda$ points. This constitutes a model-dependent indication consistent with a QGP contribution at 7.7~GeV, rather than a unique identification of QGP or a determination of its onset energy.

The comparison is qualitative: digitization does not recover complete experimental uncertainties or their correlations, and geometric centrality and decay conventions have not been fully matched to the experimental analysis. The improvement should therefore not be assigned a statistical significance from these plots alone.

\onecolumn
\twocolumn
% END ADDED APPENDICES


\begin{thebibliography}{99}
\small
\setlength{\itemsep}{0pt}
\bibitem{Fukushima2011} K. Fukushima and T. Hatsuda, Rep. Prog. Phys. \textbf{74}, 014001 (2011), doi:10.1088/0034-4885/74/1/014001.
\bibitem{Borsanyi2020} S. Bors\'anyi \textit{et al.}, Phys. Rev. Lett. \textbf{125}, 052001 (2020), doi:10.1103/PhysRevLett.125.052001.
\bibitem{Bzdak2020} A. Bzdak, S. Esumi, V. Koch, J. Liao, M. Stephanov, and N. Xu, Phys. Rep. \textbf{853}, 1 (2020), doi:10.1016/j.physrep.2020.01.005.
\bibitem{Luo2020} X. Luo, S. Shi, N. Xu, and Y. Zhang, Particles \textbf{3}, 278 (2020), doi:10.3390/particles3020022.
\bibitem{STARBESv2} L. Adamczyk \textit{et al.} (STAR Collaboration), Phys. Rev. C \textbf{88}, 014902 (2013), doi:10.1103/PhysRevC.88.014902.
\bibitem{STAR3GeV} M. S. Abdallah \textit{et al.} (STAR Collaboration), Phys. Lett. B \textbf{827}, 137003 (2022), doi:10.1016/j.physletb.2022.137003.
\bibitem{STAR3GeVErratum} STAR Collaboration, Phys. Lett. B \textbf{870}, 139912 (2025), doi:10.1016/j.physletb.2025.139912 [Erratum].
\bibitem{STARonset} B. E. Aboona \textit{et al.} (STAR Collaboration), Phys. Rev. Lett. \textbf{135}, 072301 (2025), doi:10.1103/2qhx-cp79.
\bibitem{Brachmann2000} J. Brachmann \textit{et al.}, Phys. Rev. C \textbf{61}, 024909 (2000), doi:10.1103/PhysRevC.61.024909.
\bibitem{Nara2022} Y. Nara and A. Ohnishi, Phys. Rev. C \textbf{105}, 014911 (2022), doi:10.1103/PhysRevC.105.014911.
\bibitem{RafelskiMuller1982} J. Rafelski and B. M\"uller, Phys. Rev. Lett. \textbf{48}, 1066 (1982); \textbf{56}, 2334(E) (1986), doi:10.1103/PhysRevLett.48.1066; doi:10.1103/PhysRevLett.56.2334.
\bibitem{CassingCSR} W. Cassing, A. Palmese, P. Moreau, and E. L. Bratkovskaya, Phys. Rev. C \textbf{93}, 014902 (2016), doi:10.1103/PhysRevC.93.014902.
\bibitem{Palmese2016} A. Palmese, W. Cassing, E. Seifert, T. Steinert, P. Moreau, and E. L. Bratkovskaya, Phys. Rev. C \textbf{94}, 044912 (2016), doi:10.1103/PhysRevC.94.044912.
\bibitem{Song2021} T. Song, L. Tolos, J. Wirth, J. Aichelin, and E. L. Bratkovskaya, Phys. Rev. C \textbf{103}, 044901 (2021), doi:10.1103/PhysRevC.103.044901.
\bibitem{Pal2004} S. Pal, C. M. Ko, and Z.-W. Lin, Nucl. Phys. A \textbf{730}, 143 (2004), doi:10.1016/j.nuclphysa.2003.10.013.
\bibitem{YongXi2021} G.-C. Yong, Z.-G. Xiao, Y. Gao, and Z.-W. Lin, Phys. Lett. B \textbf{820}, 136521 (2021), doi:10.1016/j.physletb.2021.136521.
\bibitem{YongQM2023} G.-C. Yong, Phys. Lett. B \textbf{843}, 138051 (2023), doi:10.1016/j.physletb.2023.138051.
\bibitem{Cassing2008} W. Cassing and E. L. Bratkovskaya, Phys. Rev. C \textbf{78}, 034919 (2008), doi:10.1103/PhysRevC.78.034919.
\bibitem{Cassing2009} W. Cassing and E. L. Bratkovskaya, Nucl. Phys. A \textbf{831}, 215 (2009), doi:10.1016/j.nuclphysa.2009.09.007.
\bibitem{Bratkovskaya2011} E. L. Bratkovskaya, W. Cassing, V. P. Konchakovski, and O. Linnyk, Nucl. Phys. A \textbf{856}, 162 (2011), doi:10.1016/j.nuclphysa.2011.03.003.
\bibitem{Moreau2019} P. Moreau, O. Soloveva, L. Oliva, T. Song, W. Cassing, and E. L. Bratkovskaya, Phys. Rev. C \textbf{100}, 014911 (2019), doi:10.1103/PhysRevC.100.014911.
\bibitem{Poskanzer1998} A. M. Poskanzer and S. A. Voloshin, Phys. Rev. C \textbf{58}, 1671 (1998), doi:10.1103/PhysRevC.58.1671.
\bibitem{STAR45} M. S. Abdallah \textit{et al.} (STAR Collaboration), Phys. Rev. C \textbf{103}, 034908 (2021), doi:10.1103/PhysRevC.103.034908.
\bibitem{E895v2} C. Pinkenburg \textit{et al.} (E895 Collaboration), Phys. Rev. Lett. \textbf{83}, 1295 (1999), doi:10.1103/PhysRevLett.83.1295.
\bibitem{Greco2003} V. Greco, C. M. Ko, and P. L\'evai, Phys. Rev. Lett. \textbf{90}, 202302 (2003), doi:10.1103/PhysRevLett.90.202302.
% BEGIN ADDED REFERENCES
\bibitem{STARFXT2026} H. Li (STAR Collaboration), EPJ Web Conf. \textbf{364}, 11001 (2026), doi:10.1051/epjconf/202636411001.
\bibitem{STARBESII2026} W. Yuan (STAR Collaboration), EPJ Web Conf. \textbf{364}, 11007 (2026), doi:10.1051/epjconf/202636411007.
% END ADDED REFERENCES
\end{thebibliography}
\end{document}